\documentclass[prd,twocolumn,showpacs,floatfix,amsmath,amssymb,floatfix]{revtex4}
\usepackage{graphicx,color,dcolumn,booktabs,bm}
\usepackage{longtable,lscape}
\usepackage{txfonts}
\usepackage{overpic}
\usepackage{amssymb}
\usepackage{indentfirst}
\usepackage{feynmf}   
\usepackage{slashed}  
\usepackage{cases}
\usepackage{color}
\usepackage{multirow}
\usepackage{epstopdf}
\usepackage{graphicx,color,dcolumn,booktabs,bm}

\graphicspath{{Figures/}} %
\usepackage[colorlinks,
                      citecolor=blue,
                      anchorcolor=red,
                      menucolor=red,
                      linkcolor=red,
                      filecolor=red,
                      runcolor=red,
                      urlcolor=blue,
                      frenchlinks=red]{hyperref}
\begin{document}
\def\pca{P_c(4380)}
\def\pcb{P_c(4450)}
\def\pcc{P_c(4312)}
\def\pcd{P_c(4440)}
\def\pce{P_c(4457)}

\title{{Exploring the molecular scenario of $P_c(4312)$, $P_c(4440)$, and $P_c(4457)$ }}

\author{Cheng-Jian Xiao$^{1}$} \email{xiaocj@ihep.ac.cn}
\author{Yin Huang$^{2}$}\email{huangy2017@buaa.edu.cn}
\author{Yu-Bing Dong$^{1,3,4}$} \email{dongyb@ihep.ac.cn}
\author{Li-Sheng Geng$^{2,5}$ \footnote{Corresponding author}} \email{lisheng.geng@buaa.edu.cn}
\author{Dian-Yong Chen$^{6}$ \footnote{Corresponding author}} \email{chendy@seu.edu.cn}
\affiliation{
$^1$Institute of High Energy Physics, Chinese Academy of Sciences,
    Beijing 100049, China\\
$^2$School of Physics and Nuclear Energy Engineering, Beihang University, Beijing 100191, China\\
$^3$ Theoretical Physics Center for Science Facilities (TPCSF), CAS, Beijing 100049, China\\
$^4$ School of Physical Sciences, University of Chinese Academy of Sciences, Beijing 101408, China \\
$^5$ International Research Center for Nuclei and Particles in the Cosmos $\&$ Beijing Key Laboratory of Advanced Nuclear Materials and Physics, Beihang University, Beijing 100191, China\\
$^6$  School of Physics, Southeast University,  Nanjing 210094, China
}

\begin{abstract}
{
In the present work, we assign the newly observed $\pcc$  as a $I(J^P)= \frac{1}{2} (\frac{1}{2})^-$ molecular state composed of $\Sigma_c \bar{D}$, while $\pcd$ and $\pce$ as $\Sigma_c \bar{D}^\ast$ molecular states with $I(J^P)= \frac{1}{2} (\frac{1}{2})^-$ and $\frac{1}{2} (\frac{3}{2})^-$, respectively. In this molecular scenario,  we investigate the $P_c\to J/\psi p$ process of these three states and further predict the ratios of the $\mathcal B(P_c\to J/\psi p)$ and those of $\mathcal B(\Lambda_b\to P_c K)$ between these three states, which could serve as a crucial test of the present molecular scenario.}
\end{abstract}


\maketitle

\section{Introduction}
\label{sec:introduction}
{
Very recently, the LHCb Collaboration reported a new narrow state $P_c(4312)$ and a two-peak structure of $P_c(4450)$ through analysing  the  data of $\Lambda_b \to  J/\psi p K $ process that was collected by the LHCb Collaboration in Run I and Run II\cite{LHCbnew,Aaij:2019vzc}. The significant of the new $P_c(4312)$ state is $7.3\sigma$ and that of the two-peak structure, which corresponding to the $P_c(4440)$ and $P_c(4457)$,  is $5.4\sigma$. The resonance parameters of three $P_c$ states are,
\begin{eqnarray}
(m,\ \Gamma)_{P_c(4312)}&=&(4311.9\pm0.7^{+6.8}_{-0.6}\,,\ 9.8\pm2.7^{+3.7}_{-4.5})\,{\rm MeV},\nonumber\\
(m,\ \Gamma)_{P_c(4440)}&=&(4440.3\pm1.3^{+4.1}_{-4.7},20.6\pm4.9^{+8.7}_{-10.1}\,)\,{\rm MeV},\nonumber\\
(m,\ \Gamma)_{P_c(4457)}&=&(4457.3\pm0.6^{+4.1}_{-1.7},6.4\pm2.0^{+5.7}_{-1.9}\,)\,{\rm MeV}.\label{eq:exp-mass-width}
\end{eqnarray}
 In addition, the LHCb Collaboration measured the  ratio $R= \mathcal{B}(\Lambda_b\to P_c K) \times \mathcal{B}(P_c \to J/\psi p)/ \mathcal{B}(\Lambda_b \to J/\psi p K)$, which are

 \begin{eqnarray}
   R_{P_c(4312)}&=&0.30 \pm 0.07^{+0.34}_{-0.09},\nonumber\\
   R_{P_c(4440)}&=&1.11 \pm 0.33^{+0.22}_{-0.10},\label{eq:exp-ratio}\\
   R_{P_c(4457)}&=&0.53 \pm 0.16^{+0.15}_{-0.13},\nonumber
 \end{eqnarray}

respectively.

This new observation is similar to but different from the analysis in 2015, where two pentaquark states, $P_c(4380) $ and $P_c(4450) $ were first reported in the $\Lambda_b \to J/\psi p K$ process \cite{Aaij:2015tga,Aaij:2016phn,Aaij:2016ymb}.
Due to a nine times larger sample of $\Lambda_b\to J/\psi pK$ than the one in 2015 , the experimentalist can perform a better analysis  nowadays. The structure $\pcb$ reported in  Ref. \cite{Aaij:2015tga} was found to be a superposition of two narrow states with a mall mass gap, which are $\pcd$ and $\pce$, while the very broad state $\pca$ was found to be insensitive to the analysis \cite{LHCbnew} and an additional narrow structure near 4.3 GeV, named $\pcc$ was observed\cite{LHCbnew}.

 The $P_c$ states were observed in the $J/\psi p$  channel and thus their quark components are more likely to be $c\bar{c} uud$, which indicates their pentaquark nature. Actually, before the observation of $P_c(4380)$ and $P_c(4450)$, there were some predictions of the hidden-charm pentaquark states\cite{Garcia-Recio:2013gaa,Molina:2012mv,Wu:2010jy,Yang:2011wz}. Stimulated by the observation of the hidden-charm pentaquark states in 2015, theorists investigated the nature of the two pentaquark states from different aspects, such as baryon-meson molecule \cite{Roca:2015dva,Chen:2015moa,Chen:2015loa,Yang:2015bmv,
Huang:2015uda,Lu:2016nnt,He:2015cea,
Eides:2015dtr,Yamaguchi:2016ote,Azizi:2016dhy,Shimizu:2016rrd}, compact pentaquark state \cite{Maiani:2015vwa,Mironov:2015ica,Anisovich:2015cia,Ghosh:2015ksa,
Ali:2016dkf,Lebed:2015tna,Li:2015gta,Wang:2015epa,
Chen:2016otp,Zhu:2015bba,Park:2017jbn,Santopinto:2016pkp,Deng:2016rus} and  kinematical effect\cite{Mikhasenko:2015vca,Guo:2015umn,Liu:2015fea}.  The studies of the $P_c(4380)$ and $P_c(4450)$ were well reviewed in Refs. \cite{ali:2017abc,espo:2017abc,chen:2017abc,lebed:2017abc,guo:2017:abc,olsen:2018abc,kar:2018abc,cerr:2018abc}.

As the pentaquark story rolls on, the new result of the $P_c$ states immediately attracted the attentions of theorists.  The authors in Refs.~\cite{Ali:2019npk,Giannuzzi:2019esi,Wang:2019got} explained the new observed $P_c$ states as compact pentaquark states in the diquark model, where the quark and diquark are the fundamental units. The analysis from constituent quark model \cite{Weng:2019ynv, Zhu:2019iwm} also supported the compact pentaqurak interpretations to the new $P_c$ states. Based on the experimental observations,  the photoproductions of the three $P_c$ states were predicted in Refs. \cite{Cao:2019kst,Wang:2019krd}. In addition,  in the vicinity of the observed $P_c$ masses there are abundant charmed meson and charmed baryon thresholds, thus, these three new observed $P_c$ states could be interpreted as  hadronic molecules.  Within the molecular scenario,  the mass spectrum ~\cite{Chen:2019bip,Chen:2019asm,He:2019ify,Liu:2019tjn,
Zhang:2019xtu,Meng:2019ilv,Mutuk:2019snd,Huang:2019jlf,
Shimizu:2019ptd,Cheng:2019obk,Guo:2019kdc,Xiao:2019aya,Guo:2019fdo, Eides:2019tgv}
and the decay properties ~\cite{Cheng:2019obk,Guo:2019kdc,Xiao:2019aya,Guo:2019fdo} were investigated by various methods.

Along the way of molecular scenario, one can find the thresholds in the mass range of new $P_c$ states  are $\Sigma^{+}_c \bar{D}^0/\Sigma_c^{++} D^-$ and $\Sigma^{+}_c \bar{D}^{\ast 0}/\Sigma_c^{++} {D}^{\ast -}$, which are 4317.73/4323.55 and 4459.75/4464.23 MeV, respectively. The mass difference between $\Sigma_c^+\bar D^0$ threshold and $P_c(4312)$ is 5.73 MeV.  While the gap between $\Sigma_c^+\bar D^{\ast0}$ threshold and  $P_c(4440)/P_c(4457)$  is $19.75/2.75$ MeV, which indicates the new $P_c$ states could be good candidates of $\Sigma_c D^{(\ast)}$ molecular states and the investigations in Ref.~\cite{Chen:2019bip,Chen:2019asm,He:2019ify,Liu:2019tjn,
Zhang:2019xtu,Meng:2019ilv,Mutuk:2019snd,Huang:2019jlf,
Shimizu:2019ptd,Cheng:2019obk,Guo:2019kdc,Xiao:2019aya,Guo:2019fdo} supports such assignment.  Considering only $S$ wave interactions,  $P_c(4312)$ can be assigned as $\Sigma_c\bar D$ molecular state with $J^P=\frac12^-$ , while $P_c(4440)$ and $P_c(4457)$ can be $\Sigma_c \bar D^\ast$ molecular states with $J^P=\frac12^-$ and $J^P=\frac32^-$, respectively. In this molecular assignment, the small mass gap of $\pcd$ and $\pce$ can result from the spin-spin interactions of the components. Similar to the case of the interactions in the quark model, the masses of the states with paralleled spins are usually a bit larger than the ones with anti-paralleled spins. However, more efforts are needed to check such  assignment. We notice that the LHCb Collaboration measured the ratios of the production fractions as shown in Eq.~(\ref{eq:exp-ratio}), which provides us an opportunity to evaluate the hadronic molecule interpretations via their decay properties,  in particular, we focus on the $J/\psi p$ mode, which is the only observed one.

This work is organized as follows. After introduction, we present the  molecular structure of the pentaquark states and relevant formulae for  the decay of $P_c \to J/\psi p$ in an effective Lagrangian approach, and in Section \ref{Sec:Num}, the numerical results and discussions are presented. Section \ref{Sec:Summary} devotes to a short summary.
}

\section{Molecular structures and decays of the $P_c$ states}
\label{Sec:Form}

\subsection{Molecular structures}
In the present work, we use an effective Lagrangian approach to describe all the involved interactions at the hadronic level. The $S$ -wave interactions between the molecular states and their components read as,

\begin{eqnarray}
&&\mathcal{L}_{P_c} =\nonumber\\&&-i g_{P_{c1}} \bar P_{c1} (x)\int dy
   \Big[ \sqrt{\frac{2}{3}} \Sigma_c^{++}(x+\omega_{\bar D\Sigma_c}y) D^{-}(x-\omega_{\Sigma_c \bar D} y) \nonumber\\
  &&\hspace{15pt}+\sqrt{\frac{1}{3 }} \Sigma_c^{+}(x+\omega_{\bar D\Sigma_c}y) \bar{D}^{0}(x-\omega_{\Sigma_c \bar D} y)  \Big] \Phi(y^2) +h.c. \nonumber\\
&&\hspace{0pt}+ g_{P_{c2}} \bar P_{c2} (x) \gamma^\mu\gamma^5\int dy
 \Big[ \sqrt{\frac23} \Sigma_c^{++}(x+\omega_{\bar D^\ast\Sigma_c}y) {D}^{\ast-}_\mu(x-\omega_{\Sigma_c \bar D^\ast} y) \nonumber\\
&&\hspace{15pt} +\sqrt{\frac13}\Sigma_c^+ (x+\omega_{\bar D^\ast\Sigma_c}y) \bar{D}^{\ast 0}_\mu(x-\omega_{\Sigma_c \bar D^\ast} y) \Big] \Phi(y^2)+h.c. \nonumber \\
&&\hspace{0pt} -i g_{P_{c3}} \bar P_{c3}^\mu (x)\int dy \Big[ \sqrt{\frac23}\Sigma_c^{++}(x+\omega_{\bar D^\ast\Sigma_c}y)
  {D}^{\ast-}_\mu(x-\omega_{\Sigma_c \bar D^\ast} y) \nonumber\\
 &&\hspace{15pt}+\sqrt{\frac13} \Sigma_c^{+}(x+\omega_{\bar D^\ast\Sigma_c}y)
  \bar{D}^{\ast 0}_\mu(x-\omega_{\Sigma_c \bar D^\ast} y) \Big] \Phi(y^2)+h.c..  \ \ \ \
  \label{Eq:Lag1}
\end{eqnarray}
where $P_{c1}$, $P_{c2}$ and $P_{c3}$ refer to $\pcc$, $\pcd$ and $\pce$, respectively, and $\omega_{ij}=m_i/(m_i+m_j)$ is a kinematical parameter with $m_i$ being the mass of the molecular components.  The correlation function $\Phi (y^2)$ is introduced to describe the distributions of the components in the  molecule, which depends only on the Jacobian coordinate $y$. The Fourier transformation of the correlation functions is,
\begin{eqnarray}
\Phi(y^2) =\int \frac{d^4 p}{(2 \pi)^4} e^{-ipy} \tilde{\Phi}(-p^2).
\end{eqnarray}

The introduced correlation function also plays the role of removing the ultraviolet divergences in  Euclidean space, which requires that the Fourier transformation of the correlation function should drop fast enough in the ultraviolet region. Generally, the Fourier transformation of the correlation function is chosen in the Gaussian form \cite{Branz:2007xp, Chen:2015igx,Faessler:2007gv, Faessler:2007us, Xiao:2016hoa},
\begin{eqnarray}
\tilde{\Phi}(-p^2) =\mathrm{Exp}\left(\frac{-p_E^2}{\Lambda^2}\right),
\end{eqnarray}
where $p_E$ is the Euclidean momentum and $\Lambda$ is the parameter which reflects the distribution of the components inside the molecular states.

\begin{figure}[hbt!]
\centering
\includegraphics[width=80mm]{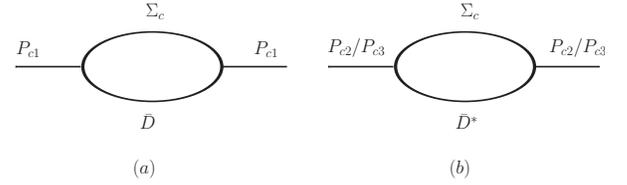}
\caption{The mass operators of the $P_{c1}$ [diagram (a)] and $P_{c2}$/$P_{c3}$ [diagram (b)], where $P_{c1}$ is assigned as a $\Sigma_c\bar{D}$ hadronic molecule with $J^P=\frac{1}{2}^-$, while $P_{c2}$ and $P_{c3}$ are $\Sigma_c \bar{D}^\ast$ hadronic molecules  with $J^P=\frac{1}{2}^-$ and $\frac32^-$ , respectively. }\label{fig:mo}
\end{figure}

The coupling constants between the hadronic molecule and its components can be determined by the compositeness condition \cite{Branz:2007xp,Weinberg:1962hj, Salam:1962ap, Hayashi:1967, Xiao:2016hoa, Chen:2015igx, Faessler:2007gv, Faessler:2007us}. For a spin-1/2 hadronic molecule, the compositeness condition is,
\begin{eqnarray}
Z\equiv1-\Sigma^\prime(m)=0,
\end{eqnarray}
where $\Sigma^\prime(m)$ is the derivative of the mass operator (as shown in Fig. \ref{fig:mo}) of the hadronic molecule. As for the spin-3/2 particle, the mass operator can be divided into the transverse and longitudinal parts, i.e.,
\begin{eqnarray}
\Sigma^{\mu\nu}(m)=g^{\mu\nu}_{\perp}
\Sigma^T(m)+\frac{p^{\mu}p^{\nu}}{p^2}
\Sigma^{L}(m).
\end{eqnarray}
And the compositeness condition for a spin-3/2 particle is,
\begin{eqnarray}
Z\equiv1-\Sigma^{T\prime}(m)=0,
\end{eqnarray}
where $\Sigma^{T\prime}(m)$ is the derivative of the  transverse part of the mass operator.

Here, the explicit form of the mass operators of $P_{c1}$, $P_{c2}$ and $P_{c3}$ are,
\begin{eqnarray}
\Sigma_{P_{c1}}(p) &=& g_{P_{c1}}^2\int
    \frac{d^4q}{(2\pi)^4} \tilde \Phi^2(q-\omega_{\Sigma\bar D}p)\frac{1}{q\!\!\!\slash - m_{\Sigma_c}}  \nonumber \\
&&\times \frac{1}{(p-q)^2-m_{\bar{D}}^2},\\
\Sigma_{P_{c2}}(p) &=& g_{P_{c2}}^2
    \int \frac{d^4q}{(2\pi)^4} \tilde \Phi^2(q-\omega_{\Sigma\bar D^\ast}p)
    \gamma^5\gamma^\mu\frac{1}{q\!\!\!\slash - m_{\Sigma_c}}
    \gamma^\mu\gamma^5  \nonumber\\
&&\times
    \frac{-g^{\mu\nu}+(p-q)^\mu(p-q)^\nu/m_{\bar D^{\ast2}}}
    {(p-q)^2-m_{\bar{D}^\ast}^2},\\
\Sigma_{P_{c3}}^{\mu\nu}(p) &=& g_{P_{c3}}^2
    \int \frac{d^4q}{(2\pi)^4} \tilde \Phi^2(q-\omega_{\Sigma\bar D^\ast}p)
    \frac{1}{q\!\!\!\slash - m_{\Sigma_c}}\nonumber\\
&&\times
    \frac{-g^{\mu\nu}+(p-q)^\mu(p-q)^\nu/m_{\bar D^{\ast2}}}
    {(p-q)^2-m_{\bar{D}^\ast}^2}.
\end{eqnarray}

\subsection{Decays of $P_{cs}' \to J/\psi p$}

Besides the effective Lagrangian presented in Eq. (\ref{Eq:Lag1}), we need additional Lagrangians related to $\Sigma_c D^{(\ast)} P$ and $ \psi D^{(\ast) } D^{(\ast)}$ interactions, which are \cite{Kaymakcalan:1983qq, Oh2000qr, Casalbuoni1996pg, Colangelo2002mj, Zou:2002yy},

\begin{eqnarray}
&&\mathcal L_{\psi D^{(*)} D^{(*)}} = -i g_{\psi D D}
    \psi_{\mu}(\partial^{\mu} D D^\dag - D \partial^{\mu} D^\dag)\nonumber\\
&&\hspace{30pt} + g_{\psi D^* D} \varepsilon^{\mu\nu\alpha\beta} \partial_{\mu}
    \psi_{\nu} (D^*_{\alpha} \overleftrightarrow \partial_{\beta} D^\dag - D \overleftrightarrow \partial_{\beta} D^{*\dag}_{\alpha})\nonumber\\
&&\hspace{30pt} + i g_{\psi D^* D^*} \psi^{\mu} (D^*_{\nu} \overleftrightarrow\partial^{\nu} D^{*\dag}_{\mu}
    + D^*_{\mu}\overleftrightarrow\partial^{\nu}D^{*\dag}_{\nu}  - D^*_{\nu} \overleftrightarrow \partial_{\mu} D^{*\nu\dag}), \nonumber\\
&&\mathcal L_{\Sigma_c N D^{(*)}} = g_{\Sigma_c N D^*} \bar{N} \gamma_{\mu}
    {\vec \tau} \cdot {\vec \Sigma_c} D^{*\mu}
-i g_{\Sigma_c N D} \bar{N} \gamma_5
    {\vec \tau} \cdot {\vec \Sigma_c} D. \ \ \
\end{eqnarray}
In the heavy quark limit, the couplings constants $g_{\psi D^{(\ast)}D^{(\ast)}}$ can be related to a universal gauge coupling $g_2$ by \cite{Kaymakcalan:1983qq, Oh2000qr, Casalbuoni1996pg, Colangelo2002mj},
\begin{eqnarray}
g_{\psi DD} &=& 2g_2\sqrt{m_{\psi}}m_D, \nonumber\\
g_{\psi D^\ast D} &=& 2g_2\sqrt{m_\psi m_{D^\ast}/m_D},\nonumber\\
g_{\psi D^\ast D^\ast} &=&  2g_2\sqrt{m_{\psi}}m_{D^\ast},
\end{eqnarray}
with $g_2=\sqrt{m_\psi}/(2m_Df_\psi)$ and $f_\psi=426$\,MeV is the decay constant of $J/\psi$, which can be estimated by the dilepton partial width of $J/\psi$ \cite{Tanabashi:2018oca}. As for the coupling constants related to the baryons, we take the same values, i.e.,  $g_{\Sigma_c ND^\ast}=3.0$ and $g_{\Sigma_c ND}=2.69 $,  as those in Refs. \cite{Dong:2009tg, Garzon:2015zva}.

\begin{figure}[hbt]
\begin{tabular}{cc}
\includegraphics[scale=0.45]{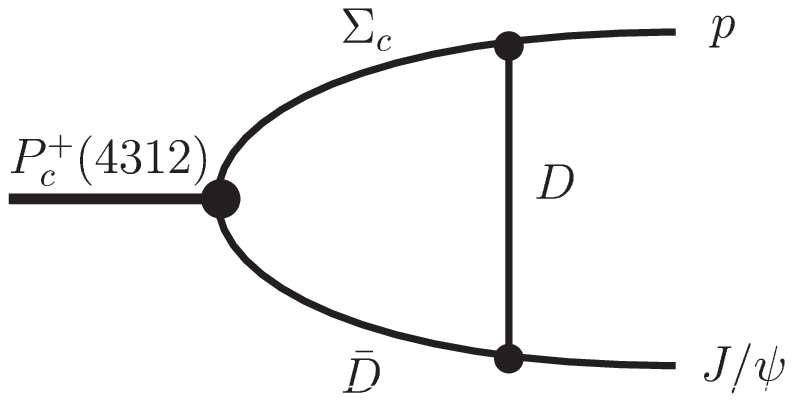}
&\includegraphics[scale=0.45]{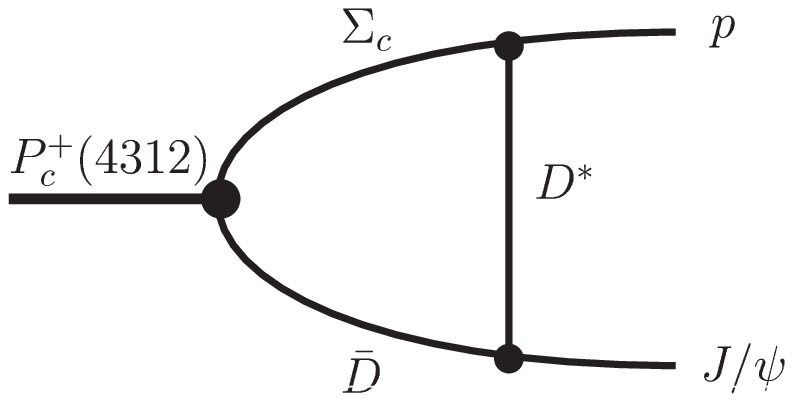}\vspace{3pt}\\
(a) &(b)\\
\\
\includegraphics[scale=0.45]{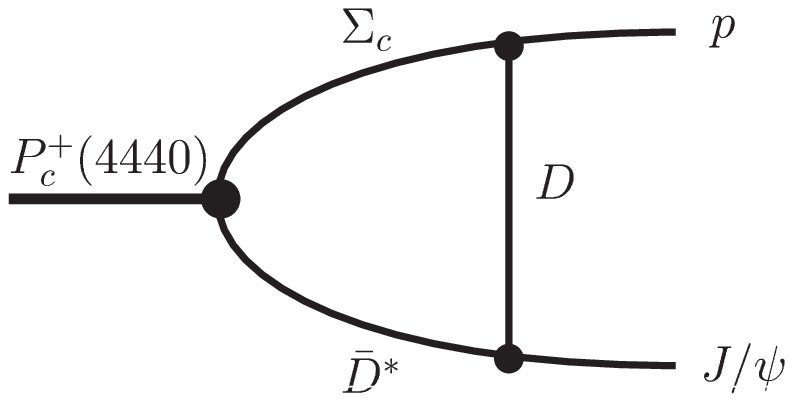}
&\includegraphics[scale=0.45]{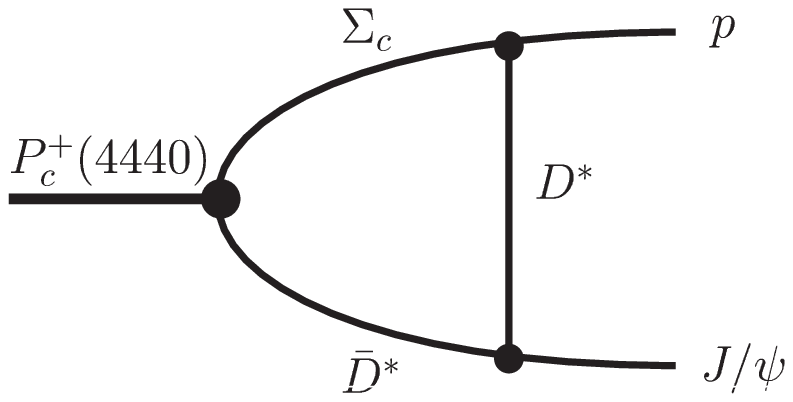}\vspace{3pt}\\
(c) &(d)\\
\end{tabular}
\caption{Feynman diagrams contributing to processes $P_{c}(4312)\to J/\psi p$ [diagram (a)-(b)] and $P_{c}(4440)\to J/\psi p$ [diagrams (c)-(d)], while the diagrams related to $P_c(4457) \to J/\psi p$ are the same as those of $P_{c}(4440)\to J/\psi p$, since the hadron components of $P_{c}(4440)$ and $P_c(4457)$ are exactly the same in the present scenario.\label{fig:tri}}
\end{figure}

In the present hadronic molecular scenario, the diagrams contributing to the $P_c\to J/\psi p$ decay are presented in Fig.~\ref{fig:tri}. In particular, for the $P_c(4312)\to J/\psi p$ decay, the amplitudes corresponding to Fig. \ref{fig:tri}-(a) -(b) are,
\begin{eqnarray}
\mathcal{M}_a&=&(i)^3\int \frac{d^4q}{(2\pi)^4}
    \tilde\Phi(-(\omega_{\bar D\Sigma_c}p_1
        -\omega_{\Sigma_c\bar D} p_2 )^2)\nonumber\\
&&\times\big[-ig_{\Sigma_cND}\bar u_{p}\gamma^5\big]
    \frac{1}{p_1\!\!\!\!\!\slash-m_1}
    \big[-ig_{P_{c1}} u_{_{P_c}}\big]\nonumber\\
&&\times\big[-ig_{\psi DD}\epsilon_\psi^\mu(-iq^\mu+ip_2^\mu)\big]\nonumber\\
&&\times
    \frac{1}{p_2^2-m_{\bar D}^2}
    \frac{1}{q^2-m_D^2},\\
\mathcal{M}_b&=&(i)^3\int \frac{d^4q}{(2\pi)^4}
    \tilde\Phi(-(\omega_{\bar D\Sigma_c}p_1
        -\omega_{\Sigma_c\bar D} p_2 )^2)\nonumber\\
&&\times\big[g_{\Sigma_cND^\ast}\bar u_p\gamma_\phi\big]
    \frac{1}{p_1\!\!\!\!\!\slash-m_1}
    \big[-ig_{P_{c1}} u_{_{P_c}}\big]\nonumber\\
&&\times
    \big[g_{\psi D^\ast D}\epsilon^{\mu\nu\alpha\beta}(ip_4^\mu)
    \epsilon^\nu_\psi(-ip_2^\beta+iq^\beta)\big]\nonumber\\
&&\times
    \frac{1}{p_2^2-m_{\bar D^\ast}^2}
   \frac{-g_{\alpha\phi}+q_{\alpha}q_{\phi}/m_{D^\ast}^2}
    {q^2-m_{D^\ast}^2}.
\end{eqnarray}
As for the $P_c(4440)\to J/\psi p$ process, the amplitudes corresponding to Fig. \ref{fig:tri}-(c)-(d) are,
\begin{eqnarray}
\mathcal{M}_c&=&(i)^3\int \frac{d^4q}{(2\pi)^4}
    \tilde\Phi(-(\omega_{\bar D^\ast\Sigma_c}p_1
        -\omega_{\Sigma_c\bar D^\ast} p_2 )^2)\nonumber\\
&&\times\big[-ig_{\Sigma_cND}\bar u_{p}\gamma^5\big]
    \frac{1}{p_1\!\!\!\!\!\slash-m_1}
    \big[g_{P_{c2}} \gamma^\phi\gamma^5 u_{_{P_c}}\big]\nonumber\\
&&\times\big[g_{\psi D^\ast D}\epsilon_{\mu\nu\alpha\beta}
        (ip_4^\mu)\epsilon^\nu_{\psi}
        (ip_2^\beta-iq^\beta)\big]\nonumber\\
&&\times
    \frac{-g_{\phi\alpha}+p_{2\phi}p_{2\alpha}/m_{\bar D^\ast}^2}{p_2^2-m_{\bar D^\ast}^2}
    \frac{1}{q^2-m_D^2},
 \end{eqnarray}
 \begin{eqnarray}
\mathcal{M}_d &=&(i)^3\int \frac{d^4q}{(2\pi)^4}
    \tilde\Phi(-(\omega_{\bar D^\ast\Sigma_c}p_1
        -\omega_{\Sigma_c\bar D^\ast} p_2 )^2)\nonumber\\
&&\times\big[g_{\Sigma_cND^\ast}\bar u_p\gamma_\mu\big]
    \frac{1}{p_1\!\!\!\!\!\slash-m_1}
    \big[g_{P_c} \gamma^\phi\gamma^5 u_{_{P_c}}\big]\nonumber\\
&&\times\big\{ig_{\psi D^\ast D^\ast}[g^{\alpha\tau}(iq^\eta-ip_2^\eta)
    +g^{\alpha\eta}(iq^\tau-ip_2^\tau)-g^{\tau\eta}\nonumber\\
&&\times(iq^\alpha-ip_2^\alpha)]\big\}
    \frac{-g_{\nu\tau}+p_{2\nu}p_{2\tau}/m_{\bar D^\ast}^2}
    {p_2^2-m_{\bar D^\ast}^2}
   \frac{-g_{\eta\mu}+q_{\eta}q_{\mu}/m_{D^\ast}^2}
    {q^2-m_{D^\ast}^2}.\nonumber\\
\end{eqnarray}

Since the components of $\pcd$ and $\pce$ are exactly the same, the diagrams contributing to $\pce \to J/\psi p$ are the same as those of $\pcd \to J/\psi p$ as shown in Fig. \ref{fig:tri}-(c)-(d). The corresponding  amplitudes are
\begin{eqnarray}
\mathcal{M}_c^{\prime}&=&(i)^3\int \frac{d^4q}{(2\pi)^4}
    \tilde\Phi(-(\omega_{\bar D^\ast\Sigma_c}p_1
        -\omega_{\Sigma_c\bar D^\ast} p_2 )^2)\nonumber\\
&&\times\big[-ig_{\Sigma_cND}\bar u_{p}\gamma^5\big]
    \frac{1}{p_1\!\!\!\!\!\slash-m_1}
    \big[-ig_{P_{c3}} u^\phi_{P_c}\big]\nonumber\\
&&\times\big[g_{\psi D^\ast D}(ip_4^\mu)\epsilon^\nu_{\psi}
        (ip_2^\beta-iq^\beta)\big]\nonumber\\
&&\times
    \frac{-g_{\phi\alpha}+p_{2\phi}p_{2\alpha}/m_{\bar D^\ast}^2}{p_2^2-m_{\bar D^\ast}^2}
    \frac{1}{q^2-m_D^2},\\
\mathcal{M}_d^{\prime}&=&(i)^3\int \frac{d^4q}{(2\pi)^4}
    \tilde\Phi(-(\omega_{\bar D^\ast\Sigma_c}p_1
        -\omega_{\Sigma_c\bar D^\ast} p_2 )^2)\nonumber\\
&&\times\big[g_{\Sigma_cND^\ast}\bar u_p\gamma_\mu\big]
    \frac{1}{p_1\!\!\!\!\!\slash-m_1}
    \big[-ig_{P_{c3}} u_{P_c}^\nu\big]\big\{ig_{\psi D^\ast D^\ast}\epsilon^\alpha_{\psi}\nonumber\\
&&\times[g^{\alpha\tau}(iq^\eta-ip_2^\eta)
    +g^{\alpha\eta}(iq^\tau-ip_2^\tau)-g^{\tau\eta}\nonumber\\
 &&\times   (iq^\alpha-ip_2^\alpha)]\big\}
    \frac{-g_{\nu\tau}+p_{2\nu}p_{2\tau}/m_{\bar D^\ast}^2}
    {p_2^2-m_{\bar D^\ast}^2}
   \frac{-g_{\eta\mu}+q_{\eta}q_{\mu}/m_{D^\ast}^2}
    {q^2-m_{D^\ast}^2}.\nonumber\\
\end{eqnarray}
With the above amplitudes, we can compute the partial decay width of $P_c \to J/\psi p$ by,
\begin{eqnarray}
\Gamma_{P_c} =\frac{1}{2J+1} \frac{1}{8\pi} \frac{|\vec p|^2}{m_0^2}  \overline{\left|  \mathcal{M}\right|^2},
\end{eqnarray}
where the $J$ is the angular momentum of the $P_c$ states and $\vec p$ is  the 3-momentum of the final states.

\section{Numerical Results and discussion}
\label{Sec:Num}

\begin{figure}[hbt]
\includegraphics[scale=0.75]{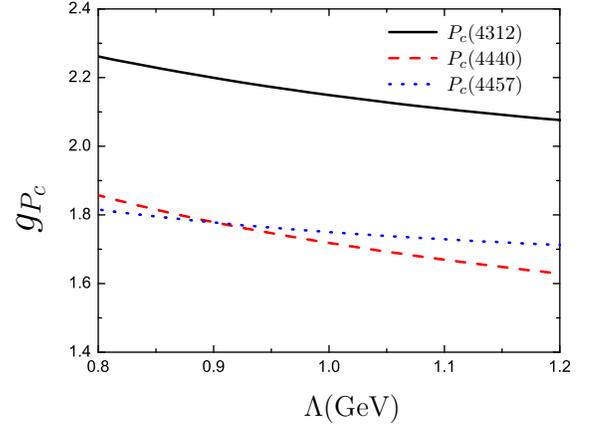}
\caption{The coupling constant $g_{P_c}$ depending on the parameter $\Lambda$. \label{fig:coup}}
\end{figure}

Before we discuss the partial decay widths of $P_c \to J/\psi p$, we need to determine the coupling constants related to the molecular state and its components. By using the compositeness condition of the molecular states, {we can estimate the coupling constants $g_{P_c}$ depending on the model parameter $\Lambda$, which is of order 1 GeV \cite{Branz:2007xp,Chen:2015igx, Faessler:2007gv, Faessler:2007us}.  However, the  accurate value of $\Lambda$ cannot be determined by the first principle. Alternatively, it is usually determined by the measured decay width.  Unfortunately,   the present experimental data is still too less to determine the $\Lambda$ for $P_c(4312)$, $P_c(4440)$ and $P_c(4457)$. Thus, in the present work, we vary $\Lambda$ from 0.8 to 1.2 GeV to check the $\Lambda$ dependence of our results.

In Fig.~\ref{fig:coup}, the $\Lambda$ dependence of the coupling constants are presented. We find that the values of the coupling constants for three $P_c$ states are very similar, especially for $P_c(4440)$ and $P_c(4457)$, which reflects the similarity of these molecular states. Moreover, the $\Lambda$ dependence of the coupling constants are similar, in particular, the coupling constants decrease with the increasing of $\Lambda$.}

\begin{figure}[hbt!]
\includegraphics[scale=0.75]{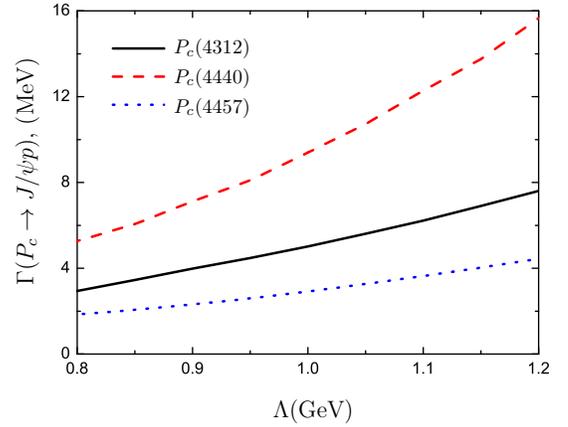}
\caption{The partial decay widths of $J/\psi p$ mode of $P_c(4312)$, $P_c(4440)$ and $P_c(4457)$, which depend on the parameter $\Lambda$.\label{fig:width} }
\end{figure}

\begin{figure}[hbt!]
\includegraphics[scale=0.7]{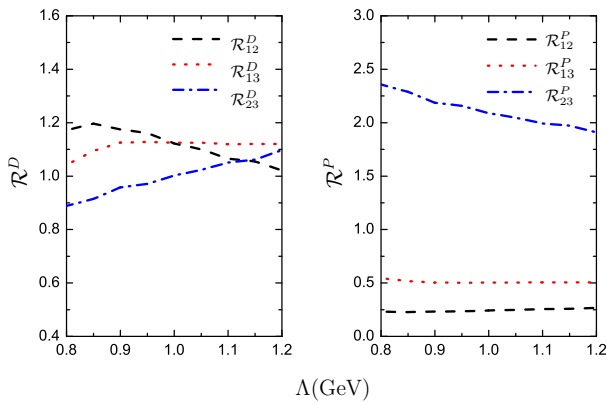}
\caption{The numerical results of decay ratios $\mathcal R^D$ in Eq.~(\ref{eq:def-RD}) (left) and production ratios in Eq.~(\ref{eq:def-RP}) (right), which depend on the parameter $\Lambda$.\label{fig:ratio}}
\end{figure}

{The estimated partial widths of $P_c \to J/\psi p$ depending on $\Lambda$ are presented in Fig.~\ref{fig:width}, where the partial widths of $P_c \to J/\psi p$ increase with the increasing of $\Lambda$.
On the one hand, our estimated results of the partial decay widths do not exceed the upper limit of the observed width, which indicates the chosen range of $\Lambda$ is reasonable. On the other hand, one may find that the estimated partial decay widths are sensitive to the $\Lambda$. Although the rough range of $\Lambda$ is determined, the accurate value of partial decay width can not be well predicted. Nevertheless, the $\pcc$, $\pcd$ and $\pce$ are considered as the molecular states composed of $\Sigma_c D^{(\ast)}$ in the present work. Both the $D$ and $D^\ast$ are $S$-wave charmed mesons and they are degenerated states in the heavy quark limit. The model parameter for $\pcc$, $\pcd$ and $\pce$ can be the same due to such similarities. Here, we define the decay ratios as,
 \begin{eqnarray}
\mathcal R^D_{12}=\mathcal{B}(P_c(4312)\to J/\psi p)/\mathcal{B}(P_c(4440)\to J/\psi p),\nonumber\\
\mathcal R^D_{13}=\mathcal{B}(P_c(4312)\to J/\psi p)/\mathcal{B}(P_c(4457)\to J/\psi p),\label{eq:def-RD}\\
\mathcal R^D_{23}=\mathcal{B}(P_c(4440)\to J/\psi p)/\mathcal{B}(P_c(4457)\to J/\psi p).\nonumber
 \end{eqnarray}
The numerical results of the decay ratios $\mathcal  R^D$ are presented in Fig.~\ref{fig:ratio} (left panel), which weakly depend on the parameter $\Lambda$. In the considered $\Lambda$ range, in particular, $\mathcal R_{12}^{D}$, $\mathcal R_{13}^{D}$ and $\mathcal R_{23}^{D}$ are predicted to be $1.17\sim1.02$, $1.04\sim1.12$ and $0.89\sim1.10$, where the central values of the observed widths were adapted in the present estimation. Since the LHCb Collaboration has measured the $R=\mathcal{B}(\Lambda_b\to P_c K) \times \mathcal{B}(P_c \to J/\psi p)/ \mathcal{B}(\Lambda_b \to J/\psi p K)$ as listed in Eq.~(\ref{eq:exp-ratio}), we can further calculate the production ratios as,
 \begin{eqnarray}
\mathcal R^P_{12}=\mathcal{B}(\Lambda_b\to P_c(4312)K)/\mathcal{B}(\Lambda_b\to P_c(4440)K),\nonumber\\
\mathcal R^P_{13}=\mathcal{B}(\Lambda_b\to P_c(4312)K)/\mathcal{B}(\Lambda_b\to P_c(4457)K),\label{eq:def-RP}\\
\mathcal R^P_{23}=\mathcal{B}(\Lambda_b\to P_c(4440)K)/\mathcal{B}(\Lambda_b\to P_c(4457)K).\nonumber
 \end{eqnarray}
The numerical results are presented in Fig.~\ref{fig:ratio} (right panel). In the considered $\Lambda$ range, $\mathcal R_{12}^{P}$, $\mathcal R_{13}^{P}$ and $\mathcal R_{23}^{P}$ are predicted to be $0.23\sim0.26$, $0.54\sim0.50$ and $2.36\sim1.91$. These predicted ratios in Eqs.~(\ref{eq:def-RD})-(\ref{eq:def-RP}) weakly depend on the  model parameter, which could serve as a crucial test  of the molecular scenario.
 }

\section{Summary}
\label{Sec:Summary}

Inspired by the recent measurement of three pentaquark states in the $J/\psi p$ invariant mass spectrum of the $\Lambda_b \to J/\psi p K$ process and noting that the newly observed states are very close to the thresholds of $\Sigma_c \bar{D}$ and $\Sigma_c \bar{D}^\ast$, we assume that the newly observed state $\pcc$ is a $I(J^P)= \frac{1}{2} (\frac{1}{2}^-)$ molecular state composed of $\Sigma_c \bar{D}$, while $\pcd$ and $\pce$ are $\Sigma_c \bar{D}^\ast$ molecular states with $I(J^P)= \frac{1}{2} (\frac{1}{2}^-)$ and $I(J^P)= \frac{1}{2} (\frac{3}{2}^-)$, respectively. In this scenario, the small mass gap of $\pcd$ and $\pce$ originates from the spin-spin interaction of the components.

In the present molecular scenario, we investigate the decays of  $P_c \to J/\psi p$ since $J/\psi p$ mode is the only observed decay pattern of $P_c$ states. Our estimations indicate the partial widths are dependent on $\Lambda$. Moreover, We present a reliable prediction for the decay ratios $\mathcal R_{12}^{D}$, $\mathcal R_{13}^{D}$ and $\mathcal R_{23}^{D}$, which are weakly dependent on the model parameter. Together with the experimental measured product of production fraction, we can estimate production ratios $\mathcal R_{12}^{P}$, $\mathcal R_{13}^{P}$ and $\mathcal R_{23}^{P}$, which are also weakly dependent on the model parameter.

Nowadays, the LHCb Collaboration have accumulated a large data sample of $\Lambda_b\to J/\psi pK$, which makes it possible to measure the decay ratios or the production ratios. The present molecular scenario can be further tested by comparing the measured values of these two ratios with our predictions.

\section*{Acknowledgement}
This work is supported in part by the National Natural Science Foundation of China (NSFC) under Grant Nos. 11775050, 11735003 and 11475192, by the fund provided to the Sino-German CRC 110 ``Symmetries and the Emergence of Structure in QCD" project by the NSFC under Grant No.11621131001, by the Key Research Program of Frontier Sciences, CAS, Grant No. Y7292610K1, by the Fundamental Research Funds for the Central Universities, and by the China Postdoctoral Science Foundation under Grant No. 2019M650843.

\end{document}